\documentstyle[psfig,doublespace]{mn}
\onecolumn
 
\def\ltsima{$\; \buildrel < \over \sim \;$}
\def\simlt{\lower.5ex\hbox{\ltsima}}
\def\gtsima{$\; \buildrel > \over \sim \;$}
\def\simgt{\lower.5ex\hbox{\gtsima}}
\def\mum{$\mu$m}
\def\msun{M$_{\odot}$}

\title[Dark Halos of Spiral Galaxies]{Dark Halos of 
Spiral Galaxies: ISO photometry
\footnote{Based on observations with ISO, an ESA project with
instruments funded by ESA Member States (especially the PI countries:
France, Germany, the Netherlands and the United Kingdom) with the
participation of ISAS and NASA.} }

\author[G.Gilmore, M.Unavane]{
Gerard~Gilmore, M.~Unavane\\
Institute of Astronomy, University of Cambridge,
Madingley Road, Cambridge, CB3 0HA
\\
}

\begin{document}
\maketitle
\begin{abstract}
We exclude hydrogen burning stars, of any mass above the
hydrogen-burning limit and any metallicity, as significant
contributors to the massive halos deduced from rotation curves to
dominate the outer parts of spiral galaxies.  We present and analyse
images of 4 nearly edge-on bulgeless spiral galaxies (UGC711, NGC2915,
UGC12426, UGC1459) obtained with ISOCAM (The CAMera instrument on
board the Infrared Space Observatory) at 14.5$\mu$m and 6.75$\mu$m.  Our
sensitivity limit for detection of any diffuse infrared emission
associated with the dark halos in these galaxies is a few tens of
microJy per 6arcsec x 6arcsec pixel, with this limit currently set by
remaining difficulties in modelling the non-linear behaviour of the
detectors.  All four galaxies show zero detected signal from extended
non-disk emission, consistent with zero halo-like luminosity density
distribution. The 95percent upper limit on any emission, for NGC2915
in particular, allows us to exclude very low mass main sequence stars
(M $>$ 0.08 Msun), and young brown dwarfs (\simlt 1 Gyr) as
significant contributors to dark matter in galactic halos. Combining
our results with those of the Galactic microlensing surveys, which
exclude objects with M $<$ 0.01 Msun, excludes almost the
entire possible mass range of compact baryonic objects from
contributing to galactic dark matter.

\end{abstract}


\section{Introduction}

The nature of the dark matter which which is required, by conventional
dynamical analyses, to dominate the outer parts of galaxies remains
unknown. A combination of conservatism and the (very) approximate
consistency between the total masses of galactic halos and the total
baryonic mass of the Universe, $\rm \Omega_B$, means that baryonic
dark matter continues to be of potential interest ({\sl cf.}
Lynden-Bell \& Gilmore 1989; Carr 1994). The mass-to-light ratios of
galactic (dark) halos are high. Thus compact baryonic candidates are
immediately restricted to stellar remnants and low mass stars and
sub-stellar mass objects.

In the Galaxy fairly detailed studies are available.  Stellar remnants
are difficult, and probably impossible, to reconcile with
well-understood limits from luminosity and chemical element production
in their luminous precursor evolutionary state (eg Hegyi \& Olive
1986; Weidemann 1989; Gibson \& Mould 1997). Additionally, direct
searches, particularly using HDF data, have produced tight limits on
their numbers (Elson, Santiago \& Gilmore 1996; Kawaler 1998). Low
mass stars have also been constrained by direct analysis of local
(Kroupa, Tout \& Gilmore 1993; Hu et al, 1994; Kirkpatrick et al 1994;
Graff \& Freese 1996; Fuchs \& Jahreiss 1998) and more distant (HST:
Santiago, Gilmore \& Elson 1996; Gould, Flynn, \& Bahcall 1998, and
refs therein; cf Mera et al 1998a, 1998b) surveys. Nonetheless, the
total number of faint stars seen by HST remains only a few hundred,
the area surveyed in any line of sight by HST is tiny, very low mass
stars do indeed exist, and the LMC microlensing events remain to be
understood (eg Alcock et al 1997). Thus improved limits on low mass
compact baryonic objects (stars?) in galactic halos remain of
interest.

The obvious observational approach in this context is to complement
deep searches for individual stars in the Galaxy halo by surface
photometry of the integrated line-of-sight emission from all halo
components in external galaxies.  The ideal candidate is an edge-on
spiral galaxy with no bulge component.  Recent efforts here include
detection (Sackett, Morrison, Harding \& Boroson 1994) and
confirmation by several groups (see refs in Lequeux et al 1998) of an
extended stellar halo around the edge-on spiral galaxy NGC5907. A
plausible explanation in this case is however that the stellar halo is
a merger remnant, similar in significance and history to the Galactic
thick disk (Gilmore \& Wyse 1985), but unrelated to the identification
of dark matter (Fuchs 1996, Lequeux et al 1998). Similar halos are
sometimes found around other galaxies (Rauscher et al 1996; Abe et al
1998), but always at surface brightnesses such that the stars actually
detected are not relevant to the dark matter problem. Dark matter is
apparently ubiquitous (eg Casertano \& van Albada 1990), and dark.

Extension of the surface photometry to the near infrared substantially
increases sensitivity to cool populations (James \& Casali 1996; Rudy
et al 1997; see also Daly \& McLaughlin 1992). It is that approach
which we have employed here, using ISO to provide the first deep
surface photometry of galaxy halos at 7$\mu$m and 15$\mu$m.

\section{Target Selection}

Suitable candidate galaxies should be known to have dark halos (hence
have measured rotation curves), be nearly edge-on to our line of sight
(to maximise disk-halo contrast), have minimal bulges (to minimise
unrelated stellar contamination), and be isolated (to minimise
confusing dynamics). Furthermore, since the observations are to be
carried out by ISO (Infrared Space Observatory), we ideally require
galaxies with linear dimensions smaller than about 3 arcminutes, so
that the whole galaxy fits into a single image field of view. Due to
the complexities of processing ISO data, this helps to minimise any
possible systematic effects.

The galaxies selected for observation, which satisfy as closely as
possible these criteria, are UGC711, NGC2915, UGC12426 and
UGC1459. These galaxies range in recession velocity from $\sim$ 500
kms$^{-1}$ to $\sim$ 5500 kms$^{-1}$, corresponding to distances from
about 5 Mpc to about 80 Mpc.

Furthermore, in the event of a detection, a simple way to distinguish
between dark matter candidates is to observe in more than one
colour. The observations were therefore in the two most sensitive
broad band ISO filters LW2 and LW3 (centred at 6.75\mum\ and 14.5\mum\
respectively).  Recent theoretical advances in cool star, brown dwarf
and cool white dwarf spectra have shown that at these wavelengths the
spectra do not deviate excessively from black body spectra.  Work by
Baraffe \& Allard (1997) shows that low mass stars and brown dwarfs,
in the very dustiest plausible case, show a diminution of flux of only
a factor of a few, at most, in the wavelength range of interest. These
models are for stars of solar metallicity, and a dark halo is more
plausibly an ancient formation, composed of metal-poor stars. In this
sense, a blackbody model may be also be suitable. Similarly, work by
Chabrier (1997) indicates that cool white dwarfs show little deviation
from black body radiation.  The difference in temperature,
nevertheless, between low mass stars or brown dwarfs and cool white
dwarfs is sufficient to show up in the LW2-LW3 colours, being of order
0.7 dex.

Accurate determination of background levels, set here almost entirely
by zodiacal emission, is crucial.  Additionally, understanding and
quantification of ISO calibrations, and time-dependant changes in
flat-fields, etc, is still improving.  Thus, for each target field
centred on the chosen galaxy, we have an offset field, observed
immediately following the target, overlapping in area, and
independently processed.  To mitigate the effect of current
uncertainty, we further choose to adopt a robust method of error
estimation. The independently processed overlapping field gives a
reliable, processing-independent method of assessing the uncertainties
in the results.

\subsection{Selection of ISO Filters}

%http://isowww.estec.esa.nl/manuals/iso_cam/node108.html
% LW filters zodiacal background signal for PFOV = 6 arcsec
% filter Wm-2sr-1 mJy/arcsec2 ADU/sec_sodi
% LW2:   1.195e-6  6.327e-2     5.67
% LW3:   3.284e-6  3.86 e-2    27.8
% filter ADU/sec/mJy/pix
% LW2:   2.19
% LW3:   1.96

Since the aim of this experiment is to extend detection limits to
cooler objects than can be studied with ground-based telescopes, we
consider the capabilities of the most sensitive ISOCAM filters. Figure 
\ref{filtersandbbody} shows the ISOCAM system senstivity (i.e. filter
transmission $\times$ detector response) superposed on a range of
blackbody flux distributions.  The ISO LW2 and LW3 filters are centred
around 6.75\mum\ and 14.5\mum\ respectively, and provide the greatest
sensitivity to the widest range of temperatures available. The
sensitivity of the filters to very cool stellar and sub-stellar
objects is shown in figure \ref{filtersens}.  This additionally
emphasises the important restriction, that we remain insensitive to
extremely cold gas clouds. Detection of very cold baryonic dark matter
requires longer wavelength observations than are possible with ISOCAM.

\begin{figure}
%\vspace{8cm}
%\hspace{4.5cm}
%\psfig{figure=fig1.eps,width=8.6cm}
%\caption{The responses for the LW2 and LW3 filters for ISOCAM are
%shown (linear, right-hand, scale) together with blackbody curves as a
%function of frequency ($\nu$) for objects of total luminosity
%1L$_{\odot}$, at the indicated temperatures.}

\label{filtersandbbody}

FIGURE 1: The responses for the LW2 and LW3 filters for ISOCAM are
shown (linear, right-hand, scale) together with blackbody curves as a
function of frequency ($\nu$) for objects of total luminosity
1L$_{\odot}$, at the indicated temperatures.

\end{figure}

\begin{figure}
%\vspace{8cm}
%\hspace{4.5cm}
%\psfig{figure=fig2.eps,width=8.6cm,rheight=5cm} 
%\caption{The fraction of incident energy flux transmitted by the
%labelled filters, as a function of (blackbody) temperature,
%illustrating the temperature sensitivity range for this experiment.}
\label{filtersens}

FIGURE 2:  The fraction of incident energy flux transmitted by the
labelled filters, as a function of (blackbody) temperature,
illustrating the temperature sensitivity range for this experiment.

\end{figure}

\section{The observations}

   \subsection{ISOCAM observing mode}

The ISO CAM instrument (Cesarsky et al 1996) was used with 6arcsec
pixels, in a somewhat non-standard observing mode. An expected
detectable signal at a S/N ratio \simgt 10 is 1-3 mJy in 200s of
integration. We need to ensure that this noise level does decrease
with increasing integration times, and to ensure maximal reliability
of any detection of low surface brightness extended emission. The CAM
detector has a special feature, requiring significant time to reach a
steady-state response to a stable input flux, and further having a
very long decay constant back to a stable state after the flux is
removed. Although corrections for these effects are feasible in
software (see below) we ensured they were minimised by adopting
relatively long stabilisation times after each pointing prior to each
integration.  That is, we chose the tradeoff between depth and
reliability to favour reliability. As the results below illustrate,
sufficient sensitivity was also achieved.

We also need to define a local ``sky'' background, adjacent to each
observation, with several criteria. The offset field must be close to
the primary pointing in time -- to minimise system drift and changing
solar elongation -- and in direction, to minimise real gradients,
especially in zodiacal emission. The offset must however be
sufficiently far from the galaxy that any real extended emission
associated with the galaxies' halo does not appear simply as a zero
offset. The target choice, noted above, ensured these offset
conditions would be met by following each primary exposure with a
following (`concatenated') exposure of a suitable background
field. The adopted field centres are listed in table 1 below.

Thus each long integration for each field consisted of 3360 separate
short exposures, with elementary integration time 2.1seconds, for a
total on-target time of two hours per source.  The spacecraft was
dithered by 2 arcseconds after every 16 images, so that a given source
fell at all points on a regular grid covering 30$\times$28 arcseconds.
Immediately following each source, an adjacent, partially overlapping
where possible, offset `sky' area was observed in exactly the same
way.  In this way, in final processing, higher resolution can be
recovered, while individual bad pixels, and the biasing effects of the
row of pixels which is missing from the ISOCAM images, do not severely
degrade the quality of the final result.

\begin{table*}

\caption{Basic Observational Information}
\label{datatable}

\begin{tabular}{lrrcrrcr}

\hline
Field & $\alpha$ (2000.0) & $\delta$ (2000.0) & Gal. & Image & Filter
& Solar & Date \\ 
  &  h m s  &  $^\circ$ $'$ $"$  & Lat. & & & Elongation & \\
\hline
UGC711   	     &   1 08 37$\cdot$0  &  1 38 28$\cdot$0    &
-60$\cdot$9 &57500601  & LW2 & 66$^\circ$ &1997 Jun 13 \\ 
UGC711-offset    &   1 08 32$\cdot$2  &  1 36 59$\cdot$0    &
&57500602  & LW2 &   &1997 Jun 13 \\ 
UGC711 	     &   1 08 37$\cdot$0  &  1 38 28$\cdot$3    &
&21201203  & LW3 & 69$^\circ$ &1996 Jun 16 \\ 
UGC711-offset    &   1 08 32$\cdot$2  &  1 36 59$\cdot$4    &
&21201206  & LW3 &   &1996 Jun 16 \\ 
\hline
NGC2915          &   9 26 00$\cdot$0  &-76 38 30$\cdot$0    & -18$\cdot$4
&62501817  & LW2 & 94$^\circ$ &1997 Aug  2 \\ 
NGC2915-offset   &   9 23 48$\cdot$0  &-76 38 30$\cdot$0    &
&62501821  & LW2 &   &1997 Aug  2 \\ 
NGC2915          &   9 26 00$\cdot$0  &-76 38 30$\cdot$0    &
&63003409  & LW3 & 93$^\circ$ &1997 Aug  7 \\ 
NGC2915-offset   &   9 23 48$\cdot$0  &-76 38 30$\cdot$0    &
&63003413  & LW3 &   &1997 Aug  7 \\ 
\hline  
UGC12426         &  23 13 32$\cdot$3  &  6 34 07$\cdot$6    & -48$\cdot$8
&54700603  & LW2 & 65$^\circ$ &1997 May 16 \\ 
UGC12426-offset  &  23 13 38$\cdot$3  &  6 34 08$\cdot$0    &
&54700604  & LW2 &   &1997 May 16 \\ 
UGC12426 	     &  23 13 32$\cdot$3  &  6 34 07$\cdot$6    &
&21600105  & LW3 & 97$^\circ$ &1996 Jun 19 \\ 
UGC12426-offset  &  23 13 38$\cdot$3  &  6 34 07$\cdot$7    &
&21600108  & LW3 &   &1996 Jun 19 \\ 
\hline
UGC1459          &   1 59 06$\cdot$7  & 36 03 47$\cdot$1    & -24$\cdot$8 &
no-obs   & LW2 &   &  \\ 
UGC1459-offset   &   1 59 06$\cdot$7  & 36 02 18$\cdot$1    &       &
no-obs   & LW2 &   &  \\ 
UGC1459          &   1 59 06$\cdot$7  & 36 03 47$\cdot$1    &
&60102102  & LW3 & 69$^\circ$ &1997 Jul  9 \\ 
UGC1459-offset   &   1 59 06$\cdot$7  & 36 02 18$\cdot$1    &
&60102107  & LW3 &   &1997 Jul  9 \\ 
\hline
\end{tabular}

\end{table*}

   \subsection{Data Reduction}
   
The standard ISO pipeline reduction of the data uses the best
available calibration files at the time of observation and the best
available algorithms at the time of reduction. As is the case with
other long-term projects, improved calibration files and more
experience in data reduction continue to become available, so that the
pipeline-processed distributed data products do not correspond to the
bestpossible reductions. In order to benefit from the best available
calibrations and algorithms, we have not utilised the standard ISO
data products, but have reprocessed the raw data using the CAM
Interactive Analysis (CIA) software\footnote{CIA is a joint
development by the ESA Astrophysics Division and the ISOCAM Consortium
led by the ISOCAM PI, C. Cesarsky, Direction des Sciences de la
Mati\`ere, C.E.A., France.}.  The data, when converted to a useable
format, are held as a cube, with the first two dimensions representing
the individual images (210 per field in this case), and the third
dimension representing the sequence of images taken.

The reduction of the raw data requires the following steps, annotated
below:

\begin{enumerate}

\item Dark correction -- library darks are subtracted from the image to leave
data corrected for dark current. 

\item Deglitching -- cosmic ray effects are quite severe. However, the
`glitches' can be reliably identified by a combination of their
position in the image and in the time sequence of the data cube. The glitches
are removed using the CIA `particule' method, by a thresholding method
in position and time.

\item Cube reduction -- at this point, the 3360 images are condensed
into 210 images, one for each dither position, by summing the
16 exposures at each dither position.

\item Transient correction -- this is the most difficult part of the
processing chain to implement successfully. The challenge with the
ISOCAM detector is that it is highly non-linear, and has a very strong
memory effect. The time for the signal on an illuminated pixel to
decay to negligible values is longer than can be implemented into an
observing schedule. The corrections were thus assigned to
software. The method adopted here (the CIA `inversion' method,
developed by Abergel \& Desert 1997) is based on the empirical
observation that the time behaviour of each CAM pixel shows dependence
on its complete history. The linear combination of all the previous
astrophysical signal, and its decay, falling upon a given pixel can be
modelled by an exponential decay. The resulting set of linear
combinations produces, at each instant, a matrix relating the observed
signal to the expected signal, and can be solved by the inversion of
the matrix. This method produces good results for images such as the
extended galactic images we have, since it was developed and optimised
for processing of images of Galactic cirrus clouds (Abergel et al. 1996).

\item Flat field correction -- a `best' flat field is selected from a
library of flat fields, and each image in the reduced data cube is
divided by this flat field. `Best' in this sense means close in time,
and with the exposure parameters most appropriate to those used in the
observations.

\item Geometrical distortion correction -- The distortion in the image
is assumed to be linear about the centre of a CAM image, and is
corrected accordingly, using the recent calibration in CIA from Herve
Aussel. The severely vignetted edge pixels were removed.

\item Raster creation -- the default methods supplied in the CIA
package at the time of writing do not make best use of the information
available in a dithered data set such as this one. In order to use the
data as well as possible, we wrote code for a specific implementation
of the `drizzle' algorithm of Fruchter \& Hook (1996).  The `drizzle'
algorithm creates an empty output image, and `rains' down onto this
empty image the pixels from the input images, cutting the flux into
appropriate fractions according to where the input pixel falls when
transformed to the same coordinate system as the output image. The
pointing information provided in the calibration files for the ISO
images was used to specify the position and orientation of these input
images. An optimal compromise between noise and resolution can be
obtained by shrinking the footprint of input pixels.

\end{enumerate}

\subsection{The reduced ISOCAM Images}   

The processed images are shown in figures \ref{ugc711} --
\ref{ugc1459}, with optical contours taken from Digitised Sky Survey
data superimposed. For each galaxy observed, images were taken centred
on the galaxy (target field) and offset from it (a control field), in
each of the filters LW2 ($\sim$6.75\mum) and LW3 ($\sim$14.5\mum).

\begin{figure}
%\vspace{1cm}
%\hspace{4.5cm}
%\psfig{figure=fig3.gif,height=12cm}
%\caption{Four images of the UGC711 observations. The left column
%shows the target and control field for the 7$\mu$m wavelength filter
%(LW2), and the right column shows the same fields observed using the
%15$\mu$m wavelength filter (LW3). The intensity scale is given at the
%bottom of the image, as is a scale bar of length 1 arcminute. The
%field size in each case is 3.2 by 3.2 arcmin.  Note that the full
%dynamic range black-white of the intensity scale of these images is
%less than 10$\mu$Jy/arcsec$^2$ for the LW2 images, and less than
%20$\mu$Jy/arcsec$^2$ for the LW3 images.  Superimposed on all the
%images are optical contours derived from the Digitised Sky Survey.
%The very bright object to the bottom right of the target field is a
%star of magnitude V=9.3}
\label{ugc711}

FIGURE 3: Four images of the UGC711 observations. The left column
shows the target and control field for the 7$\mu$m wavelength filter
(LW2), and the right column shows the same fields observed using the
15$\mu$m wavelength filter (LW3). The intensity scale is given at the
bottom of the image, as is a scale bar of length 1 arcminute. The
field size in each case is 3.2 by 3.2 arcmin.  Note that the full
dynamic range black-white of the intensity scale of these images is
less than 10$\mu$Jy/arcsec$^2$ for the LW2 images, and less than
20$\mu$Jy/arcsec$^2$ for the LW3 images.  Superimposed on all the
images are optical contours derived from the Digitised Sky Survey.
The very bright object to the bottom right of the target field is a
star of magnitude V=9.3.
\end{figure}

\begin{figure}
%\vspace{12cm}
%\hspace{4.5cm}
%\psfig{figure=/data/cass09a/gil/iso_halos/MN/ngc2915.ps,height=10cm}
%\caption{As for figure \ref{ugc711}, but for the galaxy NGC2915.}

\label{ngc2915}

FIGURE 4:  As for figure 3, but for the galaxy NGC2915.
\end{figure}

\begin{figure}
%\vspace{12cm}
%\hspace{4.5cm}
%\psfig{figure=/data/cass09a/gil/iso_halos/MN/ugc12426.ps,height=10cm}
%\caption{As for figure \ref{ugc711}, but for the galaxy UGC12426.}

\label{ugc12426}

FIGURE 5: As for figure \ref{ugc711}, but for the galaxy UGC12426.

\end{figure}

\begin{figure}
%\vspace{12cm}
%\hspace{4.5cm}
%\psfig{figure=/data/cass09a/gil/iso_halos/MN/ugc1459.ps,height=10cm}
%\caption{As for figure \ref{ugc711}, but for the galaxy UGC1459. In this
%case there was an observation only at the longer wavelength filter LW3.}
\label{ugc1459}

FIGURE 6: As for figure 3, but for the galaxy UGC1459. In this
case there was an observation only at the longer wavelength filter LW3.
\end{figure}

\noindent 
The following points may be noted concerning the images:
\begin{enumerate}

\item The images have very different mean background levels.  The
median value of the background for the LW2 images ranges, in mJy
arcsec$^{-2}$, from 0.317 (UGC711) to 0.279 (UGC12426) to 0.067
(NGC2915), while the background for the LW3 images varies from 1.45
(UGC711) to 0.75 (UGC12426) to 0.35 (NGC2915). This range is as
expected, considering that the ecliptic latitudes of these three
galaxies range from 5$^{\circ}$ $\rightarrow$ 10$^{\circ}$
$\rightarrow$ 72$^{\circ}$.  Further details are presented in table
\ref{backlevels}.  Zodiacal light dominates the background level of
these images, but correspondingly then allows an independent check on
the ISO flux calibration. To achieve this, we use the DIRBE dataset as
a resource for measurements of zodiacal dust. We show in table
\ref{backlevels} the value of zodiacal emission averaged over all
DIRBE data at that latitude, for a solar elongation of 90$^\circ$. The
true solar elongations for the observations differ by up to 25$^\circ$
from this value, but the functional dependence on angle is small near
$\epsilon=90^\circ$. The values are in agreement, providing an
independant check on the flux calibrations adopted here.

\item The target galaxies are readily visible in these infrared images.
The peak pixels are generally 2-3 $\sigma$ above the background noise.
The LW3 image of UGC12426 (figure \ref{ugc12426}) is fairly weak, and shows 
up only faintly, as a \simlt 1.5$\sigma$ perturbation.

\item The target image for UGC711 (figure \ref{ugc711}) in the LW2
band has a  bright star (V magnitude 9.3) near the
bottom right of the field.  In our analysis this
part of the image is excised with generous margins (cf figure
\ref{excisions}).

\item Prior to further analysis, all detected sources were excised
from the images. Source removal is complete for all sources
detected with integrated flux in excess of 100$\mu$Jy. These sources
are discussed in a companion paper (Ferguson, Gilmore \& Unavane 1998).

\item The data reduction still leaves considerably more
noise than the limit set by photon statistics.  The patchiness of the
background, as well as numerous `holes', or regions of locally
correlated noise, are evident.  In this experiment our photometric
procedure involves the integration of flux in large regions of the
images, to maximise sensitivity, so that we are not sensitive to any
specific local feature in the background. Given the present state of
the flat field calibrations however, we are not yet able to reach the
potential sensitivity limit contained in the data. Future re-analyses
of these data with improved flat-field calibrations will be valuable,
hopefully allowing even tighter limits on brown dwarfs to be drawn.

\end{enumerate}

\begin{table*}
\caption{Average background levels, and distance
information. Distances $d_0$, except for NGC2915 (Meurer et al. 1996),
are estimated from the recession velocity ($v_0$) using a value of
$H_0$=70 kms$^{-1}$Mpc$^{-1}$.}

\label{backlevels}

\begin{tabular}{cccccccccc}

\hline
Galaxy   & $v_0$    & $d_0$ & Ecliptic & \multicolumn{2}{c}{Solar
Elongation} & \multicolumn{2}{c}{Measured (mJy arcsec$^{-2}$)} &
\multicolumn{2}{c}{DIRBE (mJy arcsec$^{-2}$)}  \\ 
         & kms$^{-1}$ & Mpc   & Latitude &     LW2 & LW3
&    LW2        &      LW3     &     LW2       &   LW3      \\ 
\hline
UGC711   & 1982 & 28  & -5.2  & 66 & 69 &  0.317        &    1.45 
&    0.32       &   1.02     \\  
NGC2915  &  468 & 5.3 & -71.8 & 94 & 93 &  0.067        &    0.35
&    0.12       &   0.35     \\  
UGC12426 & 4720 & 67  & +10.6 & 65 & 97 &  0.279        &    0.75
&    0.29       &   0.92     \\  
UGC1459  & 5469 & 78  & +22.4 &  -- & 69 &              &    0.73
&    0.22       &   0.69     \\  
%PGC16772 &   ?  & ?   & -34.4 & -- & -- &               &
%&    0.18       &   0.54     \\  

\hline

\end{tabular}

\end{table*}

\section{MODELLING PROJECTED GALACTIC ROTATION CURVES AND MASS PROFILES}
   
We require a simple parameterisation of the rotation curves of our
target galaxies. Since this is required only to convert our flux
limits into mass-to-light ratios, and as we see below, no halo flux
was detected, simple spherical models are the most convenient and
appropriate.

For present purposes, a galactic rotation curve can be considered to
rise from zero at the centre, reaching a constant, at value V$_\infty$, at
large radii, $R$, exterior to the luminous galaxy.  
The simplest convenient functional form for a
spherical mass halo then has density proportional to $R^{-2}$ at large
radii.  An appropriate simple two-parameter mass model which works
well is given by:
\begin{equation}
\rho(R) = \frac{\rho_0}{1+(R/R_C)^2}         
\label{eq:rho}
\end{equation}
where $R$ is the spherical coordinate, $\rho_0$ the central density,
and $R_C$ denotes a characteristic scale, the `core radius'. The
corresponding velocity profile in the plane for such a density
distribution is given by
\begin{equation}
v^2 = 4\pi G\rho_0 R_C^2 \left(1-\frac{R_C}{R} tan^{-1} \frac{R}{R_C} \right)
\label{eq:v}
\end{equation}
from which $V_\infty = 4\pi G \rho_0 R_C^2$. We can therefore
prameterise any given actual 
galactic rotation curve $V(R)$ by two  parameters $R_C$ and $\rho_0$.
We emphasise that these are simply parameters, and are not intended, or
required, to have any physical significance.

This procedure of course ignores the contribution of the identified
baryons in the galaxy to the rotation curve, and so over-estimates the
(dark) halo mass. However, for the galaxies here, and at the distances
from the centres of relevance here, this effect is unimportant. For
NGC2915 for example, the galaxy for which we derive the tightest
conclusions on the nature of the dark halo, the rotation curve
analysis of Meurer et al (1996) shows the galaxy to be completely
dark-matter dominated, with identified baryons contributing only a few
percent of the rotation curve amplitude in its very outer parts. We
have additionally been conservative in fitting the rotation curves,
all of which are still rising at the last measured data point.

\subsection{Projected halo mass density}

The spherical density distribution adopted above, when viewed from a distance
$d$ sufficiently far away ($d \gg R_C$), will be seen to have
projected surface density, 
as a function of the projected distance, $r$, from the centre of the mass
distribution, given by:
\begin{equation}
\sigma(r) = \int^{\infty}_{-\infty} \rho(\sqrt{x^2+r^2}) dx
\end{equation}
For the functional form adopted here, this becomes
\begin{equation}
\sigma(r) = \frac{\pi \rho_0 R_c^2}{(r^2+R_c^2)^\frac{1}{2}}
\end{equation}
Converting to angular measure $\theta$, which represents the distance
from the centre of mass, we have
\begin{equation}
\sigma(\theta) = \frac{\pi \rho_0 R_c^2 d^2}{(d^2 \theta^2+R_c^2)^\frac{1}{2}}
\label{eq:sigma}
\end{equation}
where $d$ is the distance to the galaxy, $\theta \ll 1$, and $\sigma$ is 
measured in M$_{\odot}$sr$^{-1}$. It is this function which must be
fit to our observations, given $R_C$ and $\rho_0$ for each galaxy.

\subsection{Rotation curves}

Rotation curves for the galaxies here are available from the
literature for UGC711 (an optical rotation curve from 
Karachentsev, 1991), and for NGC2915 (HI rotation curve from Meurer et al. 1996).
For the two remaining galaxies, UGC12426 and UGC1459, we have obtained
new optical rotation curves, in collaboration with  J. Lewis. 
These rotation curves, together with the fits to the two parameters,
$R_C$ and $\rho_0$, are presented in figure \ref{rotcurves}.

\begin{figure}
%\hspace{4.5cm}
%\psfig{figure=fig7.eps,width=8.6cm}
%\caption{Rotation curves for the four galaxies observed by ISO.
%Superposed are minimised $\chi^2$ models fitted for the two parameters
%$R_C$ and $\rho_0$ as defined in equation \ref{eq:sigma}. 
%The scatter in the data for UGC711 are such that
%its parameters are clearly uncertain.}
\label{rotcurves}

FIGURE 7: Rotation curves for the four galaxies observed by ISO.
Superposed are minimised $\chi^2$ models fitted for the two parameters
$R_C$ and $\rho_0$ as defined in equation \ref{eq:sigma}. 
The scatter in the data for UGC711 are such that
its parameters are clearly uncertain.

\end{figure}

\section{ANALYSIS OF THE ISO IMAGES}

The methodology involved in defining the flux, or an upper limit on
the flux, associated with the galaxy dark halos, is well defined. By
construction, if `dark' matter halos generate the rotation curves
observed for these galaxies, then the spatial distribution of any
luminosity associated with that dark matter is known.  While a variety
of halo models can readily be attempted, the firm null results we
obtain, together with the complexities in determining the true
statistical distribution function of errors in the data, suggest the
simplest and most robust approach.  The simplest useful experiment is
to proceed exactly as is standard for determination of luminous
profiles of galaxies, and directly to sum the flux in annuli centred
at the centre of the luminous matter distribution. 

It is necessary for this experiment to excise bright regions of the
surrounding field (e.g. stars, other galaxies) prior to testing for the
presence of any remaining extended signal. We have thus excised areas
around all sources detectable as such in the images.
We then integrate the flux in annuli.
	  
The statistical uncertainty associated with this flux counting
exercise is itself very uncertain.  The systematic locally-correlated
photometric flat-field uncertainties in the reduction remain
poorly-defined, and cannot be expected to follow Gaussian distributions to
very high precision.  We are however able to derive an empirical
calibration of this uncertainty using our adjacent, but offset,
background fields.

We proceed to quantify both the sky background level, and the
associated uncertainty, by defining sets of annuli in the offset
field, exactly as in the target field. We measure the background flux
level, as a function of radius, in these annuli.  Multiple randomly
located sets of annuli yield a mean and standard deviation associated
with each of the annular sizes in the control field.  The data for
annuli centred on the target in the target field are then assigned
uncertainties equal to the standard deviations derived in the control
field.  Since the distribution function is not exactly gaussian, the
probability associated with a given `standard deviation' will not be
exactly that appropriate to a gaussian. Nonetheless, the error
distributions measured on the scales of relevance are 
close to gaussian, as can be seen in figures \ref{pic4} and \ref{pic5}
below.  We also adopt very conservative confidence limits.
In view of the (null) results below, and the rather short timescale on
which ISO data reduction and calibration is evolving, more complex
analyses are unjustified.  Finally, the photometric measurements, with
associated error bars, are tested against a dark halo model by the
minimisation of $\chi^2$ in a two-parameter model.

Shown in figure \ref{excisions} are the galaxy images, with the
regions excised indicated, and  some example annuli.
 
\begin{figure}
%\hspace{4.5cm}
%\psfig{figure=/data/cass09a/gil/iso_halos/MN/excisions.ps,width=8.6cm}
%\caption{The black regions are regions of the images excised because
%they contain detected sources.  The images are labelled with the
%galaxy and filter, while the additional label (o) refers to control
%rather than target fields. The circles, separated by 20 arcseconds in
%this figure, indicate annuli of the sort used for the flux
%integrations.}
\label{excisions}

FIGURE 8: The black regions are regions of the images excised because
they contain detected sources.  The images are labelled with the
galaxy and filter, while the additional label (o) refers to control
rather than target fields. The circles, separated by 20 arcseconds in
this figure, indicate annuli of the sort used for the flux integrations.

\end{figure}

\subsection{Quantifying flux limits from the dark halos}

We analysed above the galaxy rotation curves, to determine the two
convenient parameters $\rho_0$ and $R_c$ describing the rotation curve,
and hence the  dark halo density. We additionally have the redshift
distance $d$ for each galaxy.
We noted above that the projected luminosity distribution expected
from a simple halo model should follow
\begin{equation}
\sigma(\theta) = \frac{\pi \rho_0 R_c^2 d^2}{(d^2
\theta^2+R_c^2)^\frac{1}{2}},
\end{equation} where $\rho_0$ can be taken to be an
observed luminosity density without loss of generality. Before fitting
this functional form to our data, however, we must consider one more
practicality.

The offset between the background of the control field and of the
target field need not be perfectly zero due to gradients in zodiacal
light and galactic emission, and any uncorrected drifts in system
sensitivity. Any such offset is unimportant, since it corresponds to a
flat zero point offset, and not a gradient centred on the target
galaxy. Additionally, most offset fields overlap the galaxy field,
providing further independent checks that we are not ignoring dark
halos with very flat central luminosity profiles. The standard
procedure in IR photometry is to subtract (`beamswitch') the offset
field from the target field, providing a notionally zero background
dataset. However, it is more reliable to fit the relevant background
value directly, for each field.  Thus, we require an additional
parameter, $\sigma_0$, to handle this zero point. That is, we fit a
model of the form
\begin{equation}
\sigma(\theta) = \sigma_0 + \frac{\alpha}{(\theta^2
+\theta_0^2)^\frac{1}{2}},
\label{eq:alpha}
\end{equation} where 
$\theta$ is given in arcseconds, and $\theta_0$ is the angular
equivalent of the `core radius' $R_C$ (i.e. $\theta_0$ =$R_C / d$).
The essential astrophysics is now quantified in the parameter
$\alpha$, which is the normalisation of any luminosity associated with
a dark matter-like density profile.

The fits of this model to the data are presented in figures \ref{pic4}
and \ref{pic5}, while the derived parameters and their uncertainties
are presented in table \ref{limitstable}.  The zero point flux levels
are those presented in table \ref{backlevels} above, where they are
seen to be in agreement with DIRBE data.  Reassuringly, the best fit
model parameters, in the minimised $\chi^2$ sense, are given for four
out of the seven galaxies by (statistically insignificant) unphysical
negative values, and the other half of the time by (statistically
insignificant) positive values, in agreement with the expectation for
a zero-signal data set and approximately gaussian errors.

\begin{figure}
%\hspace{4.5cm}
%\psfig{figure=fig9.ps,width=8.6cm}
%\caption{ Results for NGC2915 and UGC711. The $\chi^2_{min}$ fits for
%projected dark halo models of the form $\alpha
%(\theta+\theta_0)^{-1/2}$, where $\theta_0$ is the apparent angular
%extent of the core radius $R_C$. A significantly non-zero value for
%$\alpha$, the slope of the solid line, would quantify any flux
%associated with emission from the dark halo matter. The best-fit model
%and its 95 per cent limits are indicated by solid and dotted lines
%respectively.  The corresponding values of $\alpha$ are listed at the
%top left of each panel. Results for the other two galaxies are
%presented in figure \ref{pic5}.}
\label{pic4}

FIGURE 9:  Results for NGC2915 and UGC711. The $\chi^2_{min}$ fits for
projected dark halo models of the form $\alpha
(\theta+\theta_0)^{-1/2}$, where $\theta_0$ is the apparent angular
extent of the core radius $R_C$. A significantly non-zero value for
$\alpha$, the slope of the solid line, would quantify any flux
associated with emission from the dark halo matter. The best-fit model
and its 95 per cent limits are indicated by solid and dotted lines
respectively.  The corresponding values of $\alpha$ are listed at the
top left of each panel. Results for the other two galaxies are
presented in figure \ref{pic5}.
\end{figure}

\begin{figure}
%\hspace{4.5cm}
%\psfig{figure=fig10.ps,width=8.6cm}
%\caption{Results for UGC12426 and UGC1459. The $\chi^2_{min}$ fits for
%projected dark halo models of the form $\alpha
%(\theta+\theta_0)^{-1/2}$, where $\theta_0$ is the apparent angular
%extent of the core radius $R_C$. A significantly non-zero value for
%$\alpha$, the slope of the solid line, would quantify any flux
%associated with emission from the dark halo matter. The best-fit model
%and its 95 per cent limits are indicated by solid and dotted lines
%respectively.  The corresponding values of $\alpha$ are given at the
%top left of each panel. Results for the other two galaxies are
%presented in figure \ref{pic4}.}
\label{pic5}

FIGURE 10: Results for UGC12426 and UGC1459. The $\chi^2_{min}$ fits for
projected dark halo models of the form $\alpha
(\theta+\theta_0)^{-1/2}$, where $\theta_0$ is the apparent angular
extent of the core radius $R_C$. A significantly non-zero value for
$\alpha$, the slope of the solid line, would quantify any flux
associated with emission from the dark halo matter. The best-fit model
and its 95 per cent limits are indicated by solid and dotted lines
respectively.  The corresponding values of $\alpha$ are given at the
top left of each panel. Results for the other two galaxies are
presented in figure \ref{pic4}.
\end{figure}

A straightforward conversion may be made from a numerical value for
the parameter $\alpha$ in equation \ref{eq:alpha} to the parameter
$\rho_0$, which has units of W Hz$^{-1}$ pc$^{-3}$ by comparison with
equation \ref{eq:sigma}.  Note however that this value of `central
luminosity density', $\rho_0$, is merely another convenient parameter
which will allow us to interpret our limits on ISO emission,
quantified through $\alpha$, in terms of real astrophysical sources of
radiation.  $\rho_0$ represents the extrapolation of a specific model
representing the outer parts of the rotation curves of these galaxies,
with deliberately no serious consideration of more realistic models of
the density law of dark matter inside the optical confines of a galaxy
(eg Hernandez \& Gilmore 1998). $\rho_0$ is {\em not} intended to
define a physical, observable, central luminosity density, but to
quantify the luminosity density associated with the outer rotation
curve in convenient units.

The eventual 95 percent confidence upper limits on the parameter
$\rho_0$ are presented in the final two columns of table
\ref{limitstable}. They represent the ISO upper limits on flux from
the dark matter which generates the rotation curves in these four
galaxies. We now proceed to their interpretation.

\begin{table*}
\caption{The derived values for the parameter $\alpha$, quantifying
the (un)detected flux associated with dark halo emission, as defined
in equation \ref{eq:alpha}, with corresponding 95 percent confidence
limits, for the four galaxies. Also listed is the conversion to the
central luminosity density parameter $\rho_0$. The formal 95\%
confidence upper limit on $\rho_0$ is separately indicated in the last
two columns.}
\label{limitstable}
\begin{tabular}{ccccccc}
Galaxy & \multicolumn{2}{c}{$\alpha\pm\Delta\alpha$} &
       \multicolumn{2}{c}{$\rho_0 \pm \Delta\rho_0$}     &
       \multicolumn{2}{c}{$\rho_0$ (upper)}\\ 
       & \multicolumn{2}{c}{($\mu$Jy arcsec$^{-2}$)} &
       \multicolumn{4}{c}{(10$^9$ W Hz$^{-1}$pc$^{-3}$)} \\ 
		& LW2  	& LW3  	& LW2  	& LW3  & LW2 & LW3\\
\hline
%UGC711	& 27.7	& 51.7	& 6.10	& 11.4 \\  %0.2203
%NGC2915	& 11.7	& 33.3	& 0.80	& 2.28 \\  %0.06841
%UGC12426	& 43.7	& 65.4	& 3.77	& 5.64 \\  %0.08625
%UGC1459	& --- 	& 26.8	& --- 	& 3.08\\   %0.114925

UGC711     & $-$1.0$\pm$8.8 & +6.7$\pm$13.3    & $-$0.2$\pm$1.9 &
+1.5$\pm$2.9    & 1.74 & 4.25 \\ %rho0=0.11 
NGC2915    & +2.5$\pm$10.2  & $-$1.1$\pm$16.6  & +0.17$\pm$0.70  &
$-$0.1$\pm$1.1 & 0.88 & 1.05 \\ %rho0=0.24 
UGC12426   & +4.3$\pm$7.4   & $-$5.3$\pm$13.8  & +0.37$\pm$0.64  &
-0.5$\pm$1.2    & 1.03 & 0.78 \\ %rho0=0.06 
UGC1459    & ---            & $-$13.6$\pm$19.6 & ---            &
-1.6$\pm$2.2    & ---  & 0.64 \\ %rho0=0.176 

\hline
\end{tabular}
\end{table*}

\section{INTERPRETATION}

We now calculate the expected fluxes which would have been measured by
ISO, for the range of models of the dark matter for which we are sensitive.

\subsection{Flux from objects -- $delta$-function masses}

We consider initially objects whose masses are distributed as a delta
function. 
Consider objects of mass $m$, with a luminosity given, as a function
of frequency, $\nu$, by $L(\nu)$, where $\int_\nu L(\nu)d\nu =
L_{tot}$.
If a halo described by the functional form of equation \ref{eq:sigma}
is composed entirely of such objects,
the flux per unit frequency reaching the observer is given by
\begin{equation}
\frac{\sigma(\theta)L(\nu)}{4\pi d^2 m} 
\end{equation}
For observations through a photometric filter, where the function
$F(\nu)$ quantifies the response as a function of frequency, the
observed flux per unit solid angle per frequency interval received at
the satelite is given by:
\begin{equation}
\frac{\rho_0 R_c^2}{4 m(d^2 \theta^2+R_c^2)^\frac{1}{2}}
\frac{\int_\nu F(\nu) L(\nu) d\nu}{\int_\nu F(\nu) d\nu}
\end{equation}
As an example, if the source spectrum is a black body spectrum, then for
objects of effective temperature $T_e$ and total luminosity $L_{tot}$,
the function $L(\nu)$ is given by:
\begin{equation}
L(\nu) = 15 L_{tot} \left( \frac{h}{\pi k T_e} \right)^4 
\frac{\nu^3}{e^{h \nu/kT_e} -1 }
\end{equation}

\subsection{Flux from objects -- mass functions}

If instead the objects are taken to be distributed with a mass
function $n(\log m)$, which represents the relative number of sources
in the range $\log m \rightarrow \log m + d(\log m)$, with lower and
upper limits given by $m_1$ and $m_2$, then the normalisation is given
by $\int_{m_1}^{m_2} n(\log m) d(\log m)$ Similarly, for a mass
function n(m), where n(m) represents the number of stars per unit mass
interval per unit volume, the spectrum of a population of this sort is
given by the weighted sum of the appropriate library of individual
spectra, $f_m(\nu)$.
\begin{equation}
  f_{tot}(\nu) = \int_{m} n(m) f_m(\nu) dm
\end{equation}

\subsubsection{Low mass main sequence stars}

We have calculated the expected flux assuming all the dark halo mass
is in normal main-sequence stars with delta-function mass functions of
mass 1.0\msun, 0.5\msun, 0.25\msun, and 0.08\msun. More plausible
models will include a mass function.  From the delta function
calculations, it is clear that only very low mass stars are viable
options. Mass functions which rise systematically towards low masses
are clearly excluded. In the interests of conservatism, we adopt a
mass function which has a maximum near 0.25\msun, and which allows for
a rise again near the hydrogen-burning limit, at 0.08\msun. For these
purposes, we adopt the observationally-derived local mass function
derived by Kirkpatrick et al. (1994).

In each case we adopt a black-body spectrum. We do this following the
spectra of Cohen et al (1995). These authors collect high-quality IR
spectra of several standard (giant -- though the differences from
dwarfs at these wavelengths are not important) stars, later used as
calibrators for ISO. It is clear from these spectra that the LW3
filter region is an excellent approximation to a black-body (ie, flat
in $\lambda^4$F$_{\lambda}$), irrespective of stellar temperature, in
all stars which lack a (thermal dust) intrinsic IR excess. In the
region of the LW2 filter, water and CO absorption becomes significant
in the coolest stars. Nonetheless, this implies a flux loss of only a
few percent in the coolest and most metal-rich stars. Metal-poor
stars, such as might plausibly make up any baryonic dark matter,
should be even better approximations to Rayleigh-Jeans spectra.  The
adopted mass function is shown in table \ref{lowmassfn}.

Table \ref{lowmassfn} also lists representative luminosities
and effective temperatures, based on a solar neigbourhood
calibration. That is, assuming the low mass stars associated with the
dark matter are metal-rich. This provides one limit on a reasonable
mass -- luminosity relation, and is the basis for models M1 to M5 below.

A halo population of exclusively low-mass stars contributing to dark
matter might be supposed to be systematically metal-poor.  The
effective temperature scale is dependant on metallicity, in a manner
which is not well calibrated.  The general effect is to associate
systematically higher temperatures with a given mass than those listed
in table \ref{lowmassfn}. Since ISO is sensitive to cool stars, this
reduces our sensitivity for this experiment.  A conservative calculation
of the amplitude of this effect is possible by adoption of the
isochrones for stars with [M/H]=$-2$ (approximately [Fe/H]=$-2.35$) and
age 10Gyr from Baraffe, Chabrier, Allard \& Hauschildt (1997). These
isochrones lie systematically hotter than the observations of
metal-poor globular cluster and field subdwarf stars at low masses
(figures  5, 6, and 8 of Baraffe et al), and so provide an upper
limit. These values are also listed in table \ref{lowmassfn}.

The range of calibrations available corresponds to a systematic change
in the expected flux in the LW2 band by a factor of about three at a
mass of 0.1\msun, rising to a reduction by a factor of about 10 at a
mass of 0.25\msun, with metal-poor stars being systematically hotter,
and hence less luminous for ISO, relative to metal-rich stars. Since
the sources are now quite warm, LW3 sensitivity is reduced below
useful limits for the hottest mass-temperature calibration at the
lowest masses. The conclusions from the LW2 limits are however
not strongly dependant on the calibration adopted. The adopted
black-body flux approximation should be more accurate for the more
metal-poor sources.

\begin{table*}
\caption{The Kirkpatrick et al. (1994) mass function for local low
mass stars.  $\xi$ is given in stars per unit mass interval per cubic
parsec. Two sets of luminosities and temperatures are given. The first
set represents metal-rich calibrations.  The final columns give the
luminosity and effective temperature scale for very metal-poor stars,
interpolated from Baraffe et al (1997:BCAH)}
\label{lowmassfn}
\begin{tabular}{cccccc}
\multicolumn{2}{c} {Solar Neighbourhood} & \multicolumn{2}{c} {Metal-Rich}
&\multicolumn{2}{c}{Metal-Poor} \\
Mass & $\xi$ (stars $M_\odot^{-1}$pc$^{-3}$) & log (L$_{bol}$/L$_\odot$) &
T$_e$ & log (L$_{bol}$/L$_\odot$) & T$_e$(BCAH) \\
\hline
0.4875 & 0.029 &  -1.15 & 4070 & -1.10 & 4580 \\
%0.4625 & 0.058 &  5.8$\times 10^{-2}$ & 3900 \\
0.4375 & 0.044 &  -1.30 & 3740 & -1.22 & 4420 \\
%0.4125 & 0.044 &  4.0$\times 10^{-2}$ & 3580 \\
0.3875 & 0.029 &  -1.50 & 3440 & -1.46 & 4290 \\
%0.3625 & 0.072 &  2.7$\times 10^{-2}$ & 3310 \\
0.3375 & 0.087 &  -1.65 & 3190 & -1.56 & 4350 \\
%0.3125 & 0.145 &  1.7$\times 10^{-2}$ & 3090 \\
0.2875 & 0.115 &  -1.85 & 3000 & -1.71 & 4150 \\
%0.2625 & 0.158 &  1.1$\times 10^{-2}$ & 2930 \\
0.2375 & 0.245 &  -2.10 & 2860 & -1.90 & 4230 \\
%0.2125 & 0.129 &  6.2$\times 10^{-3}$ & 2810 \\
0.1875 & 0.219 &  -2.35 & 2770 & -2.08 & 4060 \\
%0.1625 & 0.302 &  3.4$\times 10^{-3}$ & 2740 \\
0.1375 & 0.058 &  -2.60 & 2710 & -2.40 & 3800 \\
%0.1125 & 0.087 &  1.6$\times 10^{-3}$ & 2690 \\
0.0875 & 0.129 &  -2.95 & 2680 & -3.2: & 3000 \\
\hline
\end{tabular}
\end{table*}

\subsubsection{Cool White Dwarfs}

Some remarks concerning cool white dwarfs as major contributors to
dark halos are noted in the introduction.  Chabrier (1997) presents a
recent summary of theoretical work on cooling theory.  Masses for WDs
vary (comparatively) little (from about 0.1\msun to 1.4\msun).  For a
typical WD of mass (0.6\msun), at ages of 10 Gyr, we expect the
bolometric luminosity to be $\lg(L/L_{\odot})$ = -5.0, and
$T_{eff}$=2900K.  That is, old white dwarfs have a bolometric mass to
light ratio which a factor of approximately 100 higher than that of
stars of similar temperature, near the hydrogen burning
limit. Correspondingly, they are hard to see.  Given the high gravity,
and (probable) low metallicity, the spectra should be adequately
approximated by a black-body.  As indicated in figure
\ref{filtersens}, objects of this temperature are too warm to allow
sensitive direct detections with ISO. We quantify this in model W1
below, which is the power per unit mass in the ISO filters for a
$\delta-$function mass distribution of white dwarfs, all of mass
0.6\msun \/ and age 10Gyr.

\subsubsection{Brown Dwarfs}

Brown dwarfs are those objects of mass below the minimum mass for
hydrogen burning. Considerable progress has been made recently in
understanding their properties.
Extensive numerical work by Stevenson (1986) led to a set of scaling
relations for the luminosity and effective temperature
of brown dwarfs of given mass ($m$) and age ($t$)
(for an assumed opacity of $\kappa$=0.01), viz:
\begin{equation}
L = 1.5\times10^{-5} \left(\frac{m}{0.08M_\odot}\right)^{1.5}
                     \left(\frac{t}{5 Gyr}\right)^{-1.25}
                      L_\odot
\\
T_e = 1420 \left(\frac{m}{0.08M_\odot}\right)^{0.79}
         \left(\frac{t}{5 Gyr}\right)^{-0.31} K
\end{equation}
Analogously to the discussion of low mass hydrogen-burning stars
above, we assume that these black-body models are one limiting case,
being most appropriate for very metal-poor brown dwarfs. 

At higher metallicities, more recent theoretical work on brown dwarf
spectra has provided a more detailed view of what the mid-IR spectra
of brown dwarfs may look like. Burrows et al. (1998) use detailed
opacity models to define a brown dwarf flux model for objects of solar
metallicity. They find that, in the wavelength range 6-20 \mum, the
flux may be diminished by up to a factor of 20 compared to the
blackbody value. On the other hand, Baraffe \& Allard (1997) provide a
set of models covering the ISO filter passbands (their figure 3), and
predict deviations from a black-body spectrum which is a strong
function of temperature for metal-rich models. For brown dwarfs of
2000K, near the hydrogen-burning limit, their models suggest that the
LW2 passband is fainter than the equavalent black-body by a factor of
about two, while even at temperatures as cool as 1000K the LW2 flux
loss is only a factor of about 3-4. In both cases the LW3 flux is very
close to the black-body calculation.
That is, a simple black-body calculation, using the Stevenson scaling
relations, is sufficiently accurate, given our complete {\it a priori}
ignorance of the possible brown dwarf mass function.

We have calculated the power per unit mass appropriate to ISO
observations for four sets of models; $\delta$-function brown dwarf
mass distributions, at both high and low masses and young and old
ages, as well as power-law mass functions for both flat, i.e. equal
numbers in equal mass bins, and Salpeter-like (index $\alpha$=2.35)
mass functions, at young, intermediate and old ages, integrating over
the mass range 0.01\msun to 0.08\msun.

\section{ISO limits on low mass populations in dark halos}

We summarise in table \ref{mltable} the results of the several models
considered. The calculation in each case involved evaluation of 
\begin{equation}
\int_\nu f(\nu)
f_{LW}(\nu) d\nu / \int_\nu f_{LW}(\nu) d\nu 
\end{equation}
where $f(\nu)$ represents the spectrum of the emission, as discussed
above, and $f_{LW}(\nu)$ the filter response from figure
\ref{filtersandbbody}. The result of the calculation, for each mass
model described above, is the resulting power output per unit mass of
star through the LW2 and LW3 filters (a light-to-mass
ratio). Normalisation for the distance and halo mass density for any
galaxy then allows direct comparison with  observations.

\begin{table*}
\caption{Power per unit frequency per unit mass for different
populations in the ISO LW2 and LW3 bands, for the models described in
the text.  }
\label{mltable}
\begin{tabular}{ccccc}
Label & Source & \multicolumn{2}{c}{Value (10$^9$ W Hz$^{-1}$
 M$_{\odot}^{-1}$)} \\ 
 & & LW2 & LW3 & ratio LW2:LW3\\
\hline
M1 &     Sun		& 59.9    & 14.0 & 4.3    \\
M2 & 0.5\msun MS star	& 22.4    &  5.5  & 4.1 \\
M3 & 0.25\msun MS star  & 13.7    &  3.6  & 3.8 \\
M4 & 0.08\msun MS star  & 5.9    &  1.6  & 3.7 \\
M5 & Mass fn as in table \ref{lowmassfn} & 15.6 & 4.1 & 3.9 \\
 & & \\
B1 & 10 Gyr 0.01\msun BD &  0.03 &  0.4  & 0.07\\
B2 & 10 Gyr 0.08\msun BD &  0.2  &  0.1  & 2.4\\
B3 & 5 Gyr 0.01\msun BD &  0.1  &  0.9  & 0.2\\
B4 & 5 Gyr 0.08\msun BD &  0.4  &  0.1  & 2.8\\
B5 & 1 Gyr 0.01\msun BD &  2.7   &  4.1   & 0.7\\
B6 & 1 Gyr 0.08\msun BD &  0.9   &  0.3  & 3.5\\
 & & \\
F1 & Salpeter  10 Gyr fn & 0.2 & 0.3 & 0.6\\
F2 & Salpeter   5 Gyr fn & 0.5 & 0.6 & 0.8\\
F3 & Salpeter   1 Gyr fn & 2.7  & 1.9  & 1.5\\
F4 & Flat      10 Gyr fn & 0.3 & 0.2 & 1.4\\
F5 & Flat       5 Gyr fn & 0.5 & 0.3 & 1.8\\
F6 & Flat       1 Gyr fn & 1.7  & 0.7 & 2.5\\
 & & \\
W1 & 10 Gyr 0.6\msun  WD &  0.07 & 0.02 & 3.8  \\
 & & \\
\hline
\end{tabular}
\end{table*}

Several features of the results are immediately noticeable.
A 0.08\msun M-dwarf is two orders  of magnitude easier to see,
relative to the Sun, in the LW2 filter than in bolometric flux.
In the brown dwarf models, more power per solar mass is produced by
low mass than by higher mass young brown dwarfs, reflecting the fact that
their mass to light ratio changes by less than their mass ratio.
The reverse is true for older brown dwarfs.
The integrated power per unit solar mass for the various brown dwarf
mass functions is almost independant of the adopted mass
function. This reflects the result above: for young brown dwarfs, the
emissivity is dominated by the lowest masses, for older brown dwarfs
it is dominated by the highest masses. Thus, one has minimal
sensitivity to the mass spectrum in the range 0.01\msun to 0.08\msun.
Finally, old white dwarfs are not well suited to direct detection by
ISO. They are too hot, and have too small a surface area for their
mass, for easy detection.

\begin{figure}
%\hspace{4.5cm}
%\psfig{figure=fig11.eps,width=8.6cm}
%\caption{The emissivities of the various dark matter candidates, in
%both the LW2 and LW3 filters, are indicated vertically, for the models
%discussed in the text and presented in table \ref{mltable}. Low mass
%hydrogen-burning star models are indicated with a solid line, brown
%dwarf models with a dotted line, and the white dwarf model with a
%dashed line. The 95\% confidence limit upper limits imposed by the
%observations of each of the four galaxies is shown by a labelled thick
%solid line.}
\label{results}
FIGURE 11: The emissivities of the various dark matter candidates, in
both the LW2 and LW3 filters, are indicated vertically, for the models
discussed in the text and presented in table \ref{mltable}. Low mass
hydrogen-burning star models are indicated with a solid line, brown
dwarf models with a dotted line, and the white dwarf model with a
dashed line. The 95\% confidence limit upper limits imposed by the
observations of each of the four galaxies is shown by a labelled thick
solid line.
\end{figure}

Given these calculated central luminosity densities per unit mass,
together with the parameterised central mass densities per unit
volume, representing the rotation curve fits in figure
\ref{rotcurves}, it is straightforward to compare the models with our
observations.  The graphical comparison of these results with the
observations is given in figure \ref{results}. The solid lines
indicate the hydrogen-burning star models (M1--M4), dotted lines
represent the brown dwarf models, and the dashed line represents the
white dwarf model.  The bold lines labelled with the galaxy name
represents the results from the ISO observations, with in each case
the 95 percent upper limits indicated by the lines labelled with
the galaxy name.

Models involving significant contributions to dark halos from hydrogen
burning stars with a plausible mass function -- model M5 -- are
clearly excluded in both passbands. Even the extreme model, of a
$\delta$-function mass function of stars at the hydrogen burning limit
is directly excluded. Similarly, models involving a substantial
population of young brown dwarfs, with masses near-to, but below, 
the hydrogen-burning limit, are also directly excluded.

Interestingly, improvements in sensitivity in the longer-wavelength
LW3 data by an order or magnitude -- which might be feasible with the
present observations, given improved calibrations and data reduction
algorithms, Though may well require SIRTF, can directly test any model
involving brown dwarfs as a significant contribution to galactic dark
halos. We illustrate the present situation below, in figure
\ref{bdrange}. This shows, with better mass resolution, the emissivity
as a function of mass of brown dwarfs of a variety of ages, as they
would be observed through the ISO LW2 and LW3 filters.

\begin{figure}
%\hspace{4.5cm}
%\psfig{figure=fig12.eps,width=8.6cm}
%\caption{The flux per unit mass of various ages (1,2,5 and 10 Gyr from
%top to bottom) of brown dwarfs. The solid lines indicate observations
%with the LW2 filter and the dotted lines indicate the LW3 filter. The
%present 95\% upper limits from our observations of NGC2915 are indicated.
%}
\label{bdrange}

FIGURE 12: The flux per unit mass of various ages (1,2,5 and 10 Gyr from
top to bottom) of brown dwarfs. The solid lines indicate observations
with the LW2 filter and the dotted lines indicate the LW3 filter. The
present 95\% upper limits from our observations of NGC2915 are indicated.

\end{figure}

\section{Conclusions} 
	   
The sensitivity of these ISO observations allows us to limit the
baryonic possibilities for halo dark matter, and to exclude
hydrogen-burning stars from consideration. The ISO observations are
more sensitive and have significantly better spatial resolution than
was previously possible with observations at these wavelengths.

For all four galaxies which ISO has observed, the `best fit' models
have zero flux contributed by emission from mass distributed like the
mass which supports the rotation curve; that is, no significant
deviation from background levels is seen. By testing for the presence
of integrated dark halo-like signal, we drive limits on the maximum
amount of signal which might be present and yet not be unambiguously
detected.

Observation of two galaxies in particular, NGC2915 and UGC1459 allow
us to place useful limits on the dark matter content. From the data
for these galaxy, we find that the limit imposed by the observation
precludes the halo being composed of low mass hydrogen-burning stars;
young brown dwarfs (\simlt 1 Gyr) are also weakly ruled out. Improved
calibrations should allow a direct test of the remaining possibility
that older brown dwarfs and/or cool white dwarfs may constitute a
baryonic compact object contribution to dark halos.

\begin{figure}
%\hspace{4.5cm}
%\psfig{figure=fig13.eps,width=8.6cm}
%\caption{The solid curve shows the halo fraction upper limit (95\%
%level) as deduced by Alcock et al (1998) by combining the results of
%MACHO and EROS  (their model S, for a spherical halo). They place no
%severe constraints on more massive objects. Our upper limits (dotted
%line) limit the importance of low mass stars, leaving only the
%brown-dwarf mass range ($\sim$ 10$^{-3}$ -- 10$^{-1}$ \msun) as
%even marginally viable.  }
\label{limits}

FIGURE 13: The solid curve shows the halo fraction upper limit (95\%
level) as deduced by Alcock et al (1998) by combining the results of
MACHO and EROS  (their model S, for a spherical halo). They place no
severe constraints on more massive objects. Our upper limits (dotted
line) limit the importance of low mass stars, leaving only the
brown-dwarf mass range ($\sim$ 10$^{-3}$ -- 10$^{-1}$ \msun) as
even marginally viable.

\end{figure}

We note that the present data do not provide useful new constraints on
an extended cD-like stellar halo around these galaxies. Such a halo,
with a spatial distribution tracing that of the dark matter, though
itself forming an insignificant part of the halo matter, is a natural
expectation in standard hierachical merging models. Loosely bound
dwarf `galaxies' dispersing very early will generate free-floating
baryonic matter, which being stellar is non-dissipative, and so will
trace the dominant non-dissipative dark matter through all later
dynamical processes.  Such a process is a natural explanation of the
faint extended luminous halo around NGC5907, and should in fact be
common. Inspection of our ISO images however, shows that a `normal'
IMF provides a too-low surface brightness for ISO to detect. Optical
data are more senstitive for that experiment.

It is interesting that observations with ISO nicely complement
microlensing studies. Extant microlensing studies already severely
constrain the mass distribution of objects present in the dark halo of
our galaxy which are of too low mass to be reliably detected here.
Alcock et al (1998) show, by combining the results of both the EROS
and MACHO experiments, that objects between 10$^{-7}$\msun \/ and
10$^{-3}$\msun \/ make up less than 25 per cent of the halo dark
matter, for a variety of halo models. The solid curve in figure
\ref{limits} shows the 95 per cent upper limit for model S (spherical
halo) of Alcock et al. (1998). The constraint is weak for objects
lower in mass than $\sim$ 10$^{-7}$ \msun, and for objects higher in
mass than $\sim$ 10$^{-3}$ -- 10$^{-2}$ \msun. It is at this high end
that our results allow us to add further constraints. The limits
imposed by the observation of NGC2915 in the LW2 wavelength range
provide upper limits as shown by the dotted line in the figure.  The
remaining mass range, where the combination of current ISO
observations and microlensing surveys cannot yet place strong
limitations, is narrow: effectively the single decade in mass
encompassing the brown dwarf regime is the only mass range not yet
directly excluded as a significant contributor to baryonic dark
matter.

\end{document}